\documentstyle[preprint,aps]{revtex}
\begin{document}
\title{Effects of a magnetic field on the one-dimensional spin-orbital model}
\author{Yu-Li Lee and Yu-Wen Lee}
\address{NCTS and Physics Department, National Tsing Hua University, \\
Hsinchu, Taiwan}
\date{\today}
\maketitle

\begin{abstract}
  We study the effects of a uniform magnetic field on the one-dimensional 
spin-orbital model in terms of effective field theories. Two regions are 
examined: one around the SU(4) point (J=K/4) and the other with $K \ll J$ 
[see Eq. (1)]. We found that when $J \leq K/4$, the spin and orbital 
correlation functions exhibit power-law decay with non-universal 
exponents. In the region with $J > K/4$, the excitation spectrum has a 
gap. When the magnetic field is beyond some critical value, a quantum 
phase transition occurs. However, the correlation functions around the 
SU(4) point and the region with $K \ll J$ exhibit distinct behavior. This 
results from different structures of excitation spectra in both regime.
\end{abstract}

\pacs{PACS numbers: 75.10.Jm, 75.40.Gb}

\section{Introduction}
  
  The interest in studying the role of orbital degrees of freedom stems 
from the understanding of the magnetic structures of transition metal 
compounds\cite{Kugel82,Broh96}. In these systems, the low-lying electron 
states have orbital degeneracy as well as the usual spin degeneracy. This 
may result in interseting magnetic properties of the Mott insulating 
phase. For example, the magnetic ordering is influenced by the orbital 
structure which may change under the pressure or the magnetization is a 
nonlinear function of the magnetic field even in the case of an isotropic 
exchange interaction of the type $\vec{S}_i \cdot \vec{S}_j$\cite{Kugel82}. 
These distinguish the spin-orbital models from the ordinary Heisenberg 
model with the spin degrees of freedom only. Thus, to understand the 
magnetic properties of these compounds, an investigation of the interplay 
between spin and orbital fluctuations is necessary. 
  
  As a prototypical model in which the quantum fluctuations of the orbital 
degrees of freedom are important, we would like to consider the following 
Hamiltonian:
\begin{equation}
  H = \sum_i [K(\vec{S}_i \cdot \vec{S}_{i+1})(\vec{T}_i \cdot 
      \vec{T}_{i+1})+J_1\vec{S}_i \cdot \vec{S}_{i+1} + J_2\vec{T}_i 
      \cdot \vec{T}_{i+1}-h S^z_i],
\label{KJ}
\end{equation}
where $\vec{T}_i$ is the pseudospin to represent the orbital degrees of 
freedom. We assume that $K$ and $h$ are positive. The Hamiltonian 
(\ref{KJ}) is related to the recently discovered spin-gapped materials, 
Na$_2$Ti$_2$Sb$_2$O\cite{Axtel97} and Na$_2$V$_2$O$_5$\cite{Isob96}. 
These materials have a quasi-1D structure and are modeled by the 
quater-filled two-band Hubbard model which is equivalent to Eq. 
(\ref{KJ}) ($h=0$) in the strong Coulomb repulsion limit. Instead of a 
magnetic field, we can also impose a uniaxial pressure $P$ which gives 
rise to a term of the type $APT^z_i$ in the Hamiltonian. In other words, 
a uniaxial pressure affects the orbital sector in much the same way that 
a magnetic field affects the spin sector. Therefore, our results also 
apply to this case where the roles of spin and orbital operators 
interchange.  

  The Hamiltonian (\ref{KJ}) with $h=0$ is invariant under independent 
SU(2) rotations in the spin and orbital spaces. When $J_1=J_2=J$, it is 
also invariant under the exchange between $\vec{S}$ and $\vec{T}$. We 
restrict our consideration to this more symmetric case and comment on 
the asymmetric case in the conclusion. The phase diagram has been 
studied in Ref.\cite{Pati98,Yamash} and the properties of the ground state 
depend on the ratio $J/K$. When $J>K/4$, the ground state is doubly 
degenerate with alternating spin and orbital singlets. The excitation 
spectrum has a finite gap. In the region with $-K/4<J\leq K/4$, it is a 
critical theory. The point with $J=K/4$ is special because the symmetry is 
enlarged to SU(4)\cite{YQLi98,Yama98} and it is Bethe-ansatz 
solvable\cite{YQLi99,Suth75}. There are three gapless bosonic modes 
with the same velocity. This implies that the central charge $c=3$ and as 
shown by Affleck\cite{Affleck86}, the fixed point Hamiltonian is 
described by the SU(4)$_1$ Wess-Zumino-Novikov-Witten (WZNW) 
model. The predictions of the conformal field theory have been confirmed 
by numerical work\cite{Yama98,Frisch99}. Away from the SU(4) point, 
the low energy excitations of the Hamiltonian (\ref{KJ}) are described by 
different effective theories at different range of $J/K$. As shown in 
Ref.\cite{Azaria99}, an appropriate starting point at weak coupling 
regime ($J \gg K$) is two decoupled Heisenberg chains\cite{Nerses97} 
while near the SU(4) point, we have to begin with the O(6) Gross-Neveu 
(GN) model.  

  In the present paper, we study Eq. (\ref{KJ}) on the symmetric line with 
$h\neq 0$. The case with $J=K/4$ has been done in Ref.\cite{Yama99} 
with Bethe ansatz and numerical methods. Although the magnetic field is 
not directly coupled to the orbital degrees of freedom, it still affects the 
orbital structure. In one dimension, this is reflected on the change of the 
orbital correlation functions. For $J\leq K/4$, the excitations are gapless at 
any finite $h$ and the leading behavior of the correlation functions is the 
same as that at $J=K/4$. The orbital correlators is hardly influenced by 
the magnetic field except the $2k_F$ component. Not only the 
characteristic momentum becomes incommensurate but the corresponding 
exponent is nonuniversal and dependent of the magnetization now. 
Especially, it shows a discontinuous jump compared with that at vanishing 
magnetic field. This is because a marginally irrelevant operator at $h=0$ 
becomes a marginal one under the magnetic field. 

   For a gapped spin liquid, there is a quantum phase transition induced by 
the magnetic field. When $h>h_c$ (the critical field), the magnetization 
$M \neq 0$ whereas $M=0$ for $h<h_c$. The corresponding quantum 
critical point (QCP) is determined by the gap of the $S^z=\pm 1$ 
components. When the magnetic field exceeds this gap, the corresponding 
excitations become gapless due to the condensation of these $S^z=\pm 1$ 
bosons. For $J>K/4$ but still near the SU(4) point, the excitations with 
$S^z=0$ are still gapped and we conjecture that they are described by the 
O(4) GN model. This results in a change on the spectrum when 
$M \neq 0$. For $h>h_c$, the low energy excitations with $S^z=0$ are 
kinks while for $h<h_c$, they become massive fermions which can be 
considered as the bound states of kinks. In this case, the orbital correlators 
are exponentially decaying functions with or without algebraically 
decaying prefactors while the spin correlators show algebraic decay. 
However, when it comes to the region with $K \ll J$, the behavior 
changes though the ground states in both regions have similar properties. 
Now the quantum phase transition occurs at the value of twice the gap 
instead of the gap. In addition, both types of correlation functions exhibit 
power-law decay with universal exponents. Indeed, as suggested in 
Ref.\cite{Azaria99,Yamash}, the excitation spectrum at $K/4<J<K/2$ is 
different from that at $J\geq K/2$. In the former case, the structure factor 
has an incoherent background with the top at $q=\pi$ and a coherent 
magnon peak at $q=\pi/2$. The relative amplitude of the peak at $q=\pi/2$ 
with respect to the incoherent background at $q=\pi$ decreases with 
increasing $J/K$ and vanishes at $J=K/2$. As for the latter, there is only 
an incoherent background at $q=\pi$\cite{foot1}. Our analysis indicates 
that both regions have distinct responses to the magnetic field (or uniaxial 
pressure).

\section{The effective Hamiltonian around the SU(4) point}

   We start by a brief review of the derivation of the low energy effective 
Hamiltonian to fix our notation. Following Ref.\cite{Azaria99}, the low 
energy effective Hamiltonian of Eq. (\ref{KJ}) ($h=0$) at $J=K/4$ can 
be derived by considering the following repulsive SU(4) Hubbard model 
($U >0$) at quarter-filling:
\begin{eqnarray}
 H_U &=& -t \sum_{ia\sigma} (c^{+}_{i+1a\sigma}c_{ia\sigma}+
          {\rm H.c.}) \nonumber  \\
  & & +\frac{U}{2} \sum_{iab\sigma \sigma^{'}} n_{ia\sigma}n_{ib
             \sigma^{'}} (1-\delta_{ab}\delta_{\sigma \sigma^{'}}).
\label{Hubb}
\end{eqnarray}
Here $c^{+}_{ia\sigma}$ creates an electron with the orbital index 
$a=1,2$ and spin $\sigma =\uparrow, \downarrow$ and $n_{ia\sigma}=
c^{+}_{ia\sigma}c_{ia\sigma}$. The quarter-filling electron band 
implies that the Fermi momentum $k_{F}=\frac{\pi}{4a_0}$ where 
$a_0$ is the lattice spacing. 
 
  The low energy physics can be described by the right-moving ($R_{a
\sigma}$) and left-moving ($L_{a\sigma}$) fermions, which is related 
to the original lattice electron operator as: $c_{ia\sigma}/\sqrt{a_0}=
R_{a\sigma}(x)\exp{(ik_{F}x)}+L_{a\sigma}(x)\exp{(-ik_{F}x)}$ 
where $x=ia_0$. We introduce bosonic fields $\phi_{R(L)a\sigma}$, 
which satisfy the commutation relation $[\phi_{La\sigma}, \phi_{Rb
\sigma^{'}}]=-\frac{i}{4}\delta_{ab}\delta_{\sigma \sigma^{'}}$, to 
bosonize the right and left movers as: $R(L)_{a\sigma}=(2\pi 
a_0)^{-1/2}\eta_{a\sigma}\exp{(\pm i\sqrt{4\pi}\phi_{R(L)a\sigma})}$ 
where $\eta_{a\sigma}$ are Klein factors. It is more convenient to 
employ the following basis:
\begin{eqnarray}
 \Phi_c &=& (\phi_{1\uparrow}+\phi_{1\downarrow}+\phi_{2
            \uparrow}+\phi_{2\downarrow})/2, \nonumber  \\
 \Phi_s &=& (\phi_{1\uparrow}-\phi_{1\downarrow}+\phi_{2
            \uparrow}-\phi_{2\downarrow})/2, \nonumber  \\
 \Phi_f &=& (\phi_{1\uparrow}+\phi_{1\downarrow}-\phi_{2
            \uparrow}-\phi_{2\downarrow})/2, \nonumber  \\
 \Phi_{sf} &=& (\phi_{1\uparrow}-\phi_{1\downarrow}-\phi_{2
            \uparrow}+\phi_{2\downarrow})/2. 
\label{bos}
\end{eqnarray}
In this new basis, $\Phi_c$ represents the total charge degree of 
freedom while other bosonic fields $\Phi_a$ ($a=s, f, sf$) correspond 
to the spin-orbital degrees of freedom. 

    Substitution of  the above representations into Eq. (\ref{Hubb}) 
gives a Hamiltonian which consists of decoupled charge and 
spin-orbital degrees of freedom. The charge sector is described by a 
Gaussian model of $\Phi_c$ perturbed by an Umklapp process: 
$\cos{(\sqrt{16\pi/K_c}\Phi_c)}$ where $K_c$ is an increasing 
function of $U$ at weak coupling. As shown in Ref.\cite{Assa99}, 
there exists a critical value $U_c$ ($\sim 2.8$) such that $K_{c}=2$ 
where a Mott transition occurs and the system becomes an insulator 
when $U > U_c$. In the following, we assume that we stay in the 
insulating phase and focus on the spin-orbital sector. By introducing 
six Majorana fermions:
\begin{eqnarray}
 (\xi^1+i\xi^2)_{R(L)} &=& \frac{\eta_1}{\sqrt{\pi a_0}}\exp{(\pm i
                           \sqrt{4\pi}\Phi_{sR(L)})}, \nonumber  \\
 (\xi^3+i\xi^4)_{R(L)} &=& \frac{\eta_2}{\sqrt{\pi a_0}}\exp{(\pm i
                           \sqrt{4\pi}\Phi_{fR(L)})}, \nonumber  \\
 (\xi^5+i\xi^6)_{R(L)} &=& \frac{\eta_3}{\sqrt{\pi a_0}}\exp{(\pm i
                           \sqrt{4\pi}\Phi_{sfR(L)})},
\label{Mxi}
\end{eqnarray}
where $\eta_a$ ($a=1,2,3$) are Klein factors\cite{foot2}, The 
corresponding Hamiltonian becomes
\begin{equation}
 {\cal H} = -\frac{i}{2} \ v_f \sum_{a=1}^6 (\xi^a_R\partial_x \xi^a_R 
     -\xi^a_L\partial_x \xi^a_L)+G_3(\sum_{a=1}^6\xi^a_R\xi^a_L)^2,
\label{o6gn}
\end{equation}
where $G_3 =-Ua_0/2$. This is an O(6) GN model with marginally 
irrelevant 4-fermion interaction due to $G_3<0$. Therefore, the low 
energy behavior of the SU(4) spin-orbital model is described by the 
SO(6)$_1$ (SU(4)$_1$) WZNW model.
  
   Away from the SU(4) point, i.e. $J \neq K/4$, there is an additional
term in the Hamiltonian:
\begin{equation}
 \Delta H = \frac{G}{\pi}(\partial_x \Phi_s)^2+G(\sum_{a=3}^5 
                  \xi^a_R\xi^a_L)^2-\frac{2iG}{\pi a_0} \ \xi^6_R\xi^6_L
                  \cos{(\sqrt{4\pi}\Phi_s)},
\label{gterm}
\end{equation}
where $G=c(J-K/4)$ and $c$ is a positive constant. With the above, 
the low energy effective Hamiltonian of Eq. (\ref{KJ}) with 
$J_1=J_2$ becomes 
\begin{eqnarray}
  H &=& H_s+H_{so}+H_{int}, \nonumber  
\end{eqnarray}
where 
\begin{eqnarray}
 H_s &=& \frac{v^0_s}{2}[(\partial_x \Theta_s)^2 +(\partial_x \Phi_s)^2]+
         \frac{g_1}{\pi}(\partial_x \Phi_s)^2-\frac{h}{\sqrt{\pi}}
         \partial_x \Phi_s , \nonumber  \\
 H_{so} &=& -\frac{i}{2} \ v_f \sum_{a=3}^6 (\xi^a_R\partial_x \xi^a_R -
            \xi^a_L\partial_x \xi^a_L) +g_1(\sum_{a=3}^5\xi^a_R\xi^a_L)^2 
            +g_2\xi^6_R\xi^6_L(\sum_{a=3}^5\xi^a_R\xi^a_L), \nonumber  \\
 H_{int} &=& -\frac{i}{\pi a_0}\cos{(\sqrt{4\pi}\Phi_s)}(g_2\sum_{a=3}^5
             \xi^a_R\xi^a_L+2g_1\xi^6_R\xi^6_L),
\label{workH}
\end{eqnarray}
with $g_1=G_3+G$ and $g_2=2G_3$. Here, the velocity of $\Phi_s$ is 
represented by a different notation. Because under the magnetic field, 
the spin SU(2) rotational symmetry and the exchange symmetry are 
broken. We do not expect that all fields have the same velocity. Note 
that only $\Phi_s$ is directly coupled to $h$. Physically, this reflects the 
fact that $\Phi_s$ represents the $S^z=\pm 1$ components of magnons 
while other fields have $S^z=0$. Eq. (\ref{workH}) is our working 
Hamiltonian.

  Let us consider $H_{so}$ first. The one-loop RG equations of coupling 
constants are
\begin{equation}
  \frac{dg_{\pm}}{dl} = \frac{g^2_{\pm}}{\pi},
\label{rgeq}
\end{equation}
where $g_{\pm}=g_1\pm g_2/2$ and $l$ is the logarithm of the length 
scale. The parameter space is divided into three regions where the 
boundaries are the same as those in the case without the magnetic 
field\cite{Azaria99}: (i) Region I ($G\leq 0$). Both $g_{\pm}$ flow to 
zero couplings. (ii) Region II ($0<G<-2G_3$). $g_+$ flows to zero 
coupling while $g_-$ flows to strong coupling. (iii) Region III 
($G>-2G_3$). Both flow to strong couplings. Based on the RG flow, 
we discuss the properties of the Hamiltonian (\ref{workH}) in the 
following sections.

\section{Correlation functions around the SU(4) point}

  We now discuss the spectrum of the Hamiltonian (\ref{workH}) and the 
behavior of correlation functions around the SU(4) point.

\subsection{$J\leq K/4$}
 
 Eq. (\ref{rgeq}) tells us that $H_{so}$ describes four free massless 
Majorana fermions at low energy when $J\leq K/4$. Integrating out the 
high energy modes of these fermions generates two kinds of interactions 
through $H_{int}$: $(\partial_x \Phi_s)^2$ and $\cos{(\sqrt{16\pi}l
\Phi_s)}$ where $l$ is an integer. (It also renormalizes $v^0_s$.) The 
former can be absorbed into the $g_1$ term of $H_s$. In addition, it 
consists of higher powers of $g_1$ and $g_2$ and we do not expect this 
will change the sign of $g_1$. The latter (the cosine term) is associated 
with the lattice translation symmetry of the original lattice model: $\Phi_s 
\rightarrow \Phi_s+\sqrt{\pi}/2$ and $\xi^a_R\xi^a_L \rightarrow 
-\xi^a_R\xi^a_L$ ($a=3,\cdots,6$). They are irrelevant operators. 
Therefore, to calculate the long distance behavior of correlation 
functions, we can treat $H_s$ and $H_{so}$ as independent sectors and 
throw away $H_{int}$.

 Keeping the above approximation in mind, $H_s$ can be diagonalized as 
the following:
\begin{equation}
 H_s = \frac{v_s}{2}\left[\frac{1}{\alpha}(\partial_x \Theta_s)^2+\alpha 
       (\partial_x \Phi_s)^2\right],
\label{DHS}
\end{equation}
where $\alpha=\sqrt{1+\frac{2g_1}{\pi v^0_s}}$ and $v_s=\alpha v^0_s$. 
We also translate $\Phi_s$ as: $\Phi_s \rightarrow \Phi_s+\frac{h}
{\sqrt{\pi}v_s\alpha}x$. With the definition of the magnetization 
$M=2S^z_{tot}/N$ ($0\leq M \leq 1$) where $N$ is the number of lattice 
points, we obtain $M=\frac{2ha_0}{\pi v_s\alpha}$. (This relation is valid 
when $M \ll 1$.) In general, we expect $\alpha$ and $v_s$ depend on the 
magnetization or magnetic field. Moreover, $\alpha<1$ because $g_1<0$ 
in this case.
   
 Now we can calculate the spin and orbital correlation functions. The 
spin and orbital pseudospin operators are, respectively, defined as: 
$\vec{S}_i=\frac{1}{2}\sum_a c^+_{ia}\vec{\sigma}c_{ia}$ and $\vec{T}_i
=\frac{1}{2}\sum_\sigma c^+_{i\sigma}\vec{\tau}c_{i\sigma}$ where 
$\vec{\sigma}$ and $\vec{\tau}$ are Pauli matrices in the spin and 
orbital spaces, respectively. Near the SU(4) point, they can be 
expressed by the WZNW fields as: $\vec{S}_i/a_0=\vec{J}_s(x)+(e^{2ik_Fx}
\vec{N}_s(x)+{\rm H.c.})+(-1)^{\frac{x}{a_0}}\vec{n}_s(x)$ and 
$\vec{T}_i/a_0=\vec{J}_t(x)+(e^{2ik_Fx}\vec{N}_t(x)+{\rm H.c.})+
(-1)^{\frac{x}{a_0}}\vec{n}_t(x)$ where $x=ia_0$. We leave the detail 
expressions of these WZNW fields in Appendix {\bf A}. With the help of Eq. 
(\ref{ST}) and shifting $\Phi_s \rightarrow \Phi_s+\frac{\sqrt{\pi}M}{2a_0}x$, 
the correlation functions in the presence of a weak magnetic field can be 
shown as follows: 
\begin{eqnarray}
 \langle T^x_i(\tau)T^x_j(0) \rangle &=& \langle T^y_i(\tau)T^y_j(0) 
          \rangle = \langle T^z_i(\tau)T^z_j(0) \rangle \nonumber  \\
    &=& \frac{a_0^2}{4\pi^2}\left[\frac{1}{(v_f\tau +ix)^2}+\frac{1}
        {(v_f\tau -ix)^2}\right] \nonumber  \\
    & & +A_1\frac{\{\cos{[\frac{\pi}{2a_0}(1+M)x]}+\cos{[\frac{\pi}{2a_0}
        (1-M)x]}\}}{(v^2_s\tau^2+x^2)^{1/4\alpha}(v^2_f\tau^2+x^2)^{1/2}} 
        \nonumber  \\
    & & +(-1)^{\frac{x}{a_0}}\frac{A_2}{v_f^2\tau^2+x^2}, \nonumber  \\ 
 \langle S^z_i(\tau)S^z_j(0) \rangle &=& \frac{M^2}{4}+\frac{a_0^2}
             {4\pi^2\alpha}\left[\frac{1}{(v_s\tau +ix)^2}+\frac{1}
             {(v_s\tau -ix)^2}\right] \nonumber  \\
     & & +A_1\frac{\{\cos{[\frac{\pi}{2a_0}(1+M)x]}+\cos{[\frac{\pi}{2a_0}
         (1-M)x]}\}}{(v^2_s\tau^2+x^2)^{1/4\alpha}(v^2_f\tau^2+
         x^2)^{1/2}} \nonumber  \\
     & & +A_3\frac{\cos{[\frac{\pi}{a_0}(1-M)x]}}{(v^2_s\tau^2+
         x^2)^{1/\alpha}}, \nonumber  \\
 \langle S^+_i(\tau)S^-_j(0) \rangle &=& \frac{B_1}{(v^2_s\tau^2 +x^2
             )^{\gamma_1/2}}\left[e^{i\frac{\pi M}{2a_0}x}\left(
             \frac{v_s\tau -ix}{v_f\tau +ix}\right)+e^{-i\frac{\pi M}
             {2a_0}x}\left(\frac{v_s\tau +ix}{v_f\tau -ix}\right)\right] 
             \nonumber \\
     & & +\frac{B_2\cos{(\frac{\pi x}{2a_0})}}{(v^2_s\tau^2 +x^2
         )^{\alpha/4}(v^2_f\tau^2 +x^2)^{1/2}} \nonumber  \\
     & & +(-1)^{\frac{x}{a_0}}\frac{B_3}{(v^2_s\tau^2 +x^2
             )^{\gamma_1/2}}\left[e^{i\frac{\pi M}{2a_0}x}\left(
             \frac{v_s\tau -ix}{v_f\tau-ix}\right)+e^{-i\frac{\pi M}
            {2a_0}x}\left(\frac{v_s\tau +ix}{v_f\tau +ix}\right)\right],
\label{stcf}
\end{eqnarray}
where $\tau$ is the imaginary time and $x=a_0|i-j|$. $\gamma_1=
1+
\frac{\alpha}{2}+\frac{1}{2\alpha}>2$. $A_i$ and $B_i$ ($i=1,2,3$) are 
nonuniversal constants.

   The various characteristic momenta in Eq. (\ref{stcf}) can be 
understood as the combination of the shifted Fermi momenta under the 
magnetic field: $k_{F\uparrow}= \frac{\pi}{4a_0}(1+M)$ and 
$k_{F\downarrow}=\frac{\pi}{4a_0}(1-M)$. The structures of orbital 
(spin) correlation functions are not affected by the magnetic field 
(uniaxial pressure) except the $2k_F$ component. The original degenerate 
$2k_F$-excitations now split into two incommensurate soft modes at 
$2k_{F\uparrow}=\frac{\pi}{2a_0}(1+M)$ and $2k_{F\downarrow}= \frac{\pi}
{2a_0}(1-M)$. Moreover, the corresponding exponent $\gamma_{2k_F}=1+
\frac{1}{2\alpha} >1.5$ even at $M=0^+$ and $J=K/4$. ($\gamma_{2k_F}=1.5$ 
when $M=0$ and $J=K/4$.) This is because the marginally irrelevant 
coupling $G_3$ in Eq. (\ref{o6gn}) now turns into a marginal one and thus 
changes the compactification radius of $\Phi_s$ as shown in Eq. 
(\ref{DHS}). The $4k_{F\downarrow}$ mode does not appear in $\langle 
T^{\alpha}_i(\tau)T^{\beta}_j(0)\rangle$. (Note that $4k_{F\downarrow} 
\equiv 4k_{F\uparrow}$ (mod. $2\pi$) and thus the $4k_{F\uparrow}$ 
mode is not independent.) This means that it does not carry nontrivial 
orbital quantum numbers. The corresponding exponent $\gamma_{4k_F}=
2/\alpha =4(\gamma_{2k_F}-1)$, which is consistent with the prediction of 
conformal field theory\cite{Yama99}. The only enhanced fluctuation is the 
$2k_F$ component of $\langle S^+_i(\tau)S^-_j(0)\rangle$ (the $B_2$ term) 
of which the exponent becomes $1+\alpha/2 < 1.5$. Another interesting 
observation is that the position of the cusp at $q=\pi/2$ in the static 
transverse spin structure factor does not shift under the magnetic field.

\subsection{$J>K/4$}
 
 When $G>0$, there are two massive regions. As suggested in 
Ref.\cite{Azaria99}, the Hamiltonian (\ref{KJ}) with $J>K/4$ falls into 
the region with $0<G<-2G_3$ in the absence of the magnetic field. We 
expect that under a weak field it is still the case which corresponds to 
the region II. In this region, because $g_+$ is marginally irrelevant and 
the O(4) symmetry is restored as $g_+=0$ ($G=-2G_3$), we conjecture that 
up to logarithmic corrections, $H_{so}$ is equivalent to the O(4) GN 
model at strong coupling regime. Next, we examine the effect of $H_{int}$ 
on $H_s$. Integrating out $\xi^a$ ($a=3,\cdots,6$) gives the interaction 
term $-\frac{m}{\pi a_0}\cos{(\sqrt{4\pi}\Phi_s)}$ where $m=i(g_2\langle 
\sum_{a=3}^5 \xi^a_R\xi^a_L\rangle +2g_1\langle \xi^6_R\xi^6_L 
\rangle)$. (The term $(\partial_x \Phi_s)^2$ is possibly generated. But 
its 
effect can be absorbed into the $g_1$ term in $H_s$.) By replacing 
$H_s$ with the following one: 
\begin{equation}
 \tilde{H}_s = \int dx \ \left\{\frac{v_s}{2}[(\partial_x \Theta_s)^2 +
           (\partial_x \Phi_s)^2]-\frac{m}{\pi a_0}\cos{(\sqrt{4\pi}
           \Phi_s)}+\frac{g_1}{\pi}(\partial_x \Phi_s)^2-\frac{h}
           {\sqrt{\pi}}\partial_x \Phi_s\right\},
\label{Hs}
\end{equation}
where $v_s$ has been substituted into $v^0_s$ to incorporate its 
renormalization effect, we assume that the low energy and long distance 
behavior of the Hamiltonian (\ref{KJ}) with $J>K/4$ and $h>h_c$ is 
approximately described by $H_{so}$ and $\tilde{H}_s$. 

     By introducing a Dirac fermion $\psi_{R(L)}$, $\tilde{H}_{s}$ can 
be fermionized as:
\begin{eqnarray}
 \tilde{H}_s &=& \int dx \ [-iv_s(\psi^+_R\partial_x\psi_R-\psi^+_L
           \partial_x\psi_L)-im(\psi^+_R\psi_L-\psi^+_L\psi_R)-h(\psi^+_R
           \psi_R+\psi^+_L\psi_L)  \nonumber  \\
            & &  +\frac{g_1}{\pi}(\psi^+_R\psi_R+\psi^+_L \psi_L)^2].
\label{Hs1}
\end{eqnarray}
We see that $h$ is equivalent to the chemical potential of fermions and 
$h=h_c=|m|$ is a quantum critical point. We are concerned with the case 
where $h>h_c$. In this case, the low energy excitations of the $\Phi_s$ 
sector are described by the following effective Hamiltonian\cite{Giam99}: 
\begin{equation}
 \tilde{H} = \int dx \ \frac{\tilde{v}_s}{2}\left[\frac{1}{g}(\partial_x 
               \tilde{\theta})^2+g(\partial_x \tilde{\phi})^2\right],
\label{Hs2}
\end{equation}
where $\tilde{\phi}$ is obtained via shifting $\Phi_s$ by an amount 
$\frac{\sqrt{\pi}M}{2a_0}x$. $g$ is a parameter determined by $g_1$ and 
$g<1$ ($g>1$) when $g_1<0$ ($g_1>0$). The scattering ($g_1$) term has 
only negligible effects in the limit $M \rightarrow 0$ where we will get free 
fermions with $g \rightarrow 1$. 

   Based on our assumption that $H_{so}$ is equivalent to an O(4) GN model 
in the strong coupling regime, we replace $H_{so}$ with the following one:
\begin{equation}
 H_{so} = -\frac{i}{2} \ v_f \sum_{a=3}^6 (\xi^a_R\partial_x \xi^a_R -
          \xi^a_L\partial_x \xi^a_L)+\tilde{g}(\sum_{a=3}^6\xi^a_R
          \xi^a_L)^2,
\label{gnm}
\end{equation}
where $\tilde{g}>0$. We have performed the duality transformation: 
$\xi^6_R \rightarrow \xi^6_R$, $\xi^6_L \rightarrow -\xi^6_L$, and 
$\sigma_6 \leftrightarrow \mu_6$. The same transformation {\it must} be 
applied to Eq. (\ref{ST}) before calculating correlation functions. The 
spectrum of O(4) GN model is distinguished from that of the O(N) GN model 
with $N\geq 6$\cite{Shan78}. The latter consists of the elementary fermions, 
the bound states of them, and kinks and can be qualitatively captured by 
the large N approximation. The existence of kinks is intimately related to 
the spontaneous breaking of discrete chiral symmetry ($Z_2$ symmetry), 
which corresponds to the breaking of lattice translation symmetry in our 
case. On the contrary, in the O(4) case, the elementary fermions are 
unstable against decay into kinks which become the only stable 
excitations. As a consequence, the magnetic field dramatically changes the 
spectrum of the $S^z=0$ sector. As shown in Ref.\cite{Azaria99}, the low 
energy excitations in the absence of the magnetic field are massive fermions 
which belong to the representations $(S_{tot}, T_{tot})=(1,0), (0,1)$. Under 
a magnetic field, the symmetry becomes U(1)$\times$SU(2) and the good 
quantum numbers are $S^z$ and $T_{tot}$. As we expect, the excitations 
of $\Phi_s$ carry $(S^z, T_{tot})=(\pm 1, 0)$. Nevertheless, those massive 
fermions carrying $S^z=0$ are no longer stable when $h>h_c$. The 
elementary excitations turns out to be kinks which carry $(S^z, T_{tot})=
(0, 1/2)$.   

   A convenient way to deal with Eq. (\ref{gnm}) is to transform it into 
two decoupled sine-Gordon models\cite{Witten78}:   
\begin{equation}
 H_{so} = \sum_{a=\pm} \left\{\frac{\tilde{v}_f}{2}\left[\frac{1}{l}
         (\partial_x\Theta_a)^2+l (\partial_x\Phi_a)^2\right]-
         \frac{\tilde{g}}{\pi^2a^2_0}\cos{\sqrt{8\pi}\Phi_a}\right\},
\label{deq}
\end{equation}
where $l=\sqrt{1+\frac{2\tilde{g}}{\pi v_f}}$ and $\tilde{v}_f=lv_f$. 
Since $l=1^+$, the coupling constant of the sine-Gordon model 
$\beta^2=8\pi^{-}$. The kinks of the O(4) GN model are nothing but 
the solitons of the two sine-Gordon models. Within our approximation 
Eqs. (\ref{Hs2}) and (\ref{gnm}), we are able to discuss the 
dynamical structure factors of spin and orbital sectors, which are 
defined as: 
\begin{eqnarray}
   S^{\alpha \beta}(\omega, q) &=& 2 \ {\rm Im} \lim_{i\omega 
       \rightarrow \omega +i0^+}\int^{\infty}_{-\infty}dx d\tau \ 
       \langle S^\alpha_i(\tau)S^\beta_j(0) \rangle e^{i\omega \tau 
       -iqx}, \nonumber  \\
   T^{\alpha \beta}(\omega, q) &=& 2 \ {\rm Im} \lim_{i\omega 
       \rightarrow \omega +i0^+}\int^{\infty}_{-\infty}dx d\tau \ 
       \langle T^\alpha_i(\tau)T^\beta_j(0) \rangle e^{i\omega \tau 
       -iqx},
\label{dsf}    
\end{eqnarray}
where $x=a_0|i-j|$. For $T^{\alpha \beta}$, it suffices to compute 
$T^{zz}$ because the orbital SU(2) symmetry remains intact. We shall see 
that the $2k_F$ components depend on $x+y$ as well as $x-y$. This is a 
manifestation of the spontaneous $Z_2$ symmetry breaking of the ground 
state. In that case, we just list the correlators in the coordinate space. 
The details of computations are left in Appendix {\bf B} and we show the 
results in the following. First, we consider the spin correlators. They 
are
\begin{eqnarray}
 S^{zz}(\omega, q) &=& \frac{a^2_0}{g\tilde{v}_s} \ \omega 
           [\delta (\omega -\tilde{v}_sq)+\delta (\omega +\tilde{v}_sq)] 
           +C_1\left(\frac{4\tilde{v}^2_s/a^2_0}{\omega^2-\tilde{v}^2_s(q
           -4k_{F\downarrow})^2}\right)^{1-\frac{1}{g}}  \nonumber  \\ 
   & & \times [\Theta(\omega -\tilde{v}_s(q-4k_{F\downarrow}))
           \Theta(\omega +\tilde{v}_s(q-4k_{F\downarrow}))-(\omega 
           \rightarrow -\omega)], \nonumber  \\
\langle S^+_i(\tau)S^-_j(0) \rangle &\sim& \cos{(\frac{\pi}{2a_0}x)}
             \cos{(\frac{\pi}{2a_0}y)}\left(\frac{a_0}{|z|}
             \right)^{\frac{g}{2}},
\label{ds1}
\end{eqnarray}
where $x=ia_0$, $y=ja_0$, and $z=\tilde{v}_s\tau +i(x-y)$. $C_1$ is a 
nonuniversal constant. 
For $S^{zz}$, the $C_1$ term is correct only near its low energy threshold 
and $q\approx 4k_{F\downarrow}$. In addition, it diverges at the low 
energy threshold as $[\omega \pm \tilde{v}_s(q-4k_{F\downarrow})
]^{-1+\frac{1}{g}}$ for $g>1$ while it approaches zero with the same 
functional form for $g<1$. As $M \rightarrow 0$, the exponent becomes 
zero. For $\langle S^+_i(\tau)S^-_j(0) \rangle$, as the case with $J\leq K/4$, 
the characteristic momentum $q=\pi/2$ does not shift under the magnetic 
field. Moreover, the exponent approaches $\frac{1}{2}$ as $M \rightarrow 
0$. 

   Next, we consider the orbital correlators. The results are as follows:
\begin{eqnarray}
 T^{zz}(\omega , q\approx 0) &=& 8a^2_0\frac{m^2v^2_fq^2
           |f(2\chi(s))|^2}{s^3\sqrt{s^2-4m^2}}sgn(\omega), \nonumber  \\
 \langle T^z_i(\tau)T^z_j(0) \rangle |_{2k_F} &\sim& 2\cos{[\frac{\pi}
           {2a_0}(x+y)]}\cos{[\frac{\pi M}{2a_0}(x-y)]}\left(\frac{a_0}
           {|z|}\right)^{\frac{1}{2g}} \nonumber  \\
   & & \times [W_1(\tau ,x-y)-W_2(\tau ,x-y)]  \nonumber \\
   & & +\{\cos{[2k_{F\uparrow}(x-y)]}+\cos{[2k_{F\downarrow}
       (x-y)]}\}\left(\frac{a_0}{|z|}\right)^{\frac{1}{2g}} \nonumber  
       \\
   & & \times [W_1(\tau ,x-y)+W_2(\tau ,x-y)], \nonumber  \\
 T^{zz}(\omega , q\approx \frac{\pi}{a_0}) &=& C_2 \frac{m^2}{u
        \sqrt{u^2-4m^2}}\times \frac{|F_0(2\chi (u))|^2}{\cos{(
        \frac{\pi^2}{\zeta})}+\cosh{[\frac{2\pi}{\zeta}\chi (u)]}}
        sgn(\omega),
\label{ds2}
\end{eqnarray}
where $x=ia_0$, $y=ja_0$, $z=\tilde{v}_s\tau+i(x-y)$ and $m$ is the 
mass of kinks. $s^2=\omega^2-v^2_fq^2$, $u^2=\omega^2-
v^2_f(q-\pi/a_0)^2$ and $\zeta=\pi \beta^2/(8\pi -\beta^2)$ with 
$\beta^2=8\pi^-$. $\chi(s)=\cosh^{-1}{(s/2m)}$. The functions $f(x)$ 
and $F_0(x)$ are defined in Eqs. (\ref{t4}) and (\ref{t7}), respectively. 
$C_2$ is a mere constant. The result for $T^{zz}(\omega, q\approx 0)$ 
is exact when $4m^2<s^2<16m^2$ while that for $T^{zz}(\omega , q
\approx \frac{\pi}{a_0})$ is valid when $4m^2 < u^2 < 16m^2$. For 
higher energy, there are small contributions from four, six, eight, etc. 
particles states. The functions $W_{1,2}$, which are defined in the 
following: 
\begin{eqnarray}
 W_1(\tau ,x) &=& F(\sin{[\sqrt{\pi/2}(\Phi_++\Phi_-)]}\sin{[\sqrt{\pi/2}
        (\Theta_+-\Theta_-)]}), \nonumber  \\
 W_2(\tau ,x) &=& F(\cos{[\sqrt{\pi/2}(\Phi_++\Phi_-)]}\cos{[\sqrt{\pi/2}
        (\Theta_+-\Theta_-)]}), 
\label{wf}
\end{eqnarray}
with $F(\hat{O}) \equiv \langle \hat{O}(\tau,x)\hat{O}(0,0) \rangle$, 
involve the computation of form factors containing $\Phi_{\pm}$ and 
their dual fields $\Theta_{\pm}$. At present, no one knows how to 
compute them. However, we can still discuss some features of these 
functions. The operators we are considering are like $\hat{A}_+ \cdot 
\hat{A}_-$ where $\hat{A}_{\pm}=\cos{[{\rm or} \ \sin{](\sqrt{\pi/2}
\Phi_{\pm})}}\times \cos{[{\rm or} \ \sin{](\sqrt{\pi/2}\Theta_{\pm})}
}$. According to Ref.\cite{Delf99}, the operators $\cos{[{\rm or} \ 
\sin{](\sqrt{\pi/2}\Theta_{\pm})}}$ are fermionic while $\cos{[{\rm or} 
\ \sin{](\sqrt{\pi/2}\Phi_{\pm})}}$ are bosonic for $\beta^2=8\pi^-$. 
Thus, $\hat{A}_{\pm}$ must be fermonic. This implies that the first 
nontrivial form factors of $\hat{A}_{\pm}$ start from the one-soliton 
states. As a result, the leading contributions to $W_i$ come from 
two-particle states. We expect that in the momentum space, the $2k_F$ 
component exhibits incoherent background with two-particle thresholds. 

    In summary, when $h>h_c$, the spin correlators exhibit algebraic 
decay while orbital correlators remain massive behavior. Low energy 
modes appear close to $q=0$ and $4k_{F\downarrow}$ for $S^{zz}$ or 
$q=\frac{\pi}{2a_0}$ for $S^{+-}$. The $2k_F$ component of $S^{zz}$ 
and the uniform and $4k_F$ components of $S^{+-}$ involve massive 
excitations [see Eq. (\ref{ST})] and thus exhibit similar behavior to 
that of $\langle T^z_i(\tau)T^z_j(0) \rangle |_{2k_F}$. The leading 
contributions to the uniform and $4k_F$ components of $\langle 
T^{\alpha}_i(\tau)T^{\beta}_j(0) \rangle$ come from the two-particle 
states, which is similar to that in the absence of the magnetic field. 
The main distinction between the cases with $h>h_c$ and $h<h_c$ is 
the $2k_F$ component. For the former case, it also exhibits two-particle 
thresholds in contrast to the latter case where it shows a coherent 
"magnon" peak. This can be detected experimentally by examining the 
dynamical spin structure factors under a uniaxial pressure.
    
\section{Weak coupling regime: $K \ll J$}

  When $K \ll J$, a proper starting point is to consider the $K$ term in 
Eq. (\ref{KJ}) as a perturbation, which was discussed by Nersesyan 
and Tsvelik\cite{Nerses97} in the context of two-leg spin ladders. The 
unperturbed Hamiltonian is composed of two antiferromagnetic (AF) 
Heisenberg chains which can be treated by using the standard relation 
between the spin (pseudospin) operators and WZNW 
fields\cite{Affleck89}: $\vec{S}_n/a_0=\vec{j}_{s}+
(-1)^{\frac{x}{a_0}}\vec{m}_s$ and $\vec{T}_n/a_0=\vec{j}_{t}+
(-1)^{\frac{x}{a_0}}\vec{m}_t$ where $x=na_0$. By defining 
$\phi_{\pm}=(\phi_s\pm \phi_t)/\sqrt{2}$ where $\phi_s$ and $\phi_t$ 
are, respectively, the bosonic fields describing the spin and orbital 
degrees of freedom, we obtain
\begin{equation}
 H = \sum_{a=\pm}\int dx \ \left\{\frac{v}{2}[(\partial_x\theta_a)^2
     +(\partial_x\phi_a)^2]+\frac{m}{\pi a_0}\cos{(\sqrt{4\pi}\phi_a)}
     -\frac{h}{\sqrt{4\pi}}\partial_x\phi_a\right\},
\label{wH1}
\end{equation}
where $v\sim Ja_0$ and $m=\lambda^2K/(2\pi^3)$. The Hamiltonian 
(\ref{wH1}) can be further fermionized as:
\begin{eqnarray}
 H &=& \sum_{a=\pm}\int dx \ [-iv(\psi^+_{aR}\partial_x\psi_{aR}-
       \psi^+_{aL}\partial_x\psi_{aL})+im(\psi^+_{aR}\psi_{aL}-
       \psi^+_{aL}\psi_{aR})  \nonumber  \\
     & & -\frac{h}{2}(\psi^+_{aR}\psi_{aR}+\psi^+_{aL}\psi_{aL})].  
\label{wH2}
\end{eqnarray}

    From Eq. (\ref{wH2}), we see that in the region with $K \ll J$, the 
low energy excitations in the absence of the magnetic field are described 
by two free massive Dirac fermions or four massive Majorana fermions, 
which are kinks connecting two degenerate 
ground states\cite{Nerses97}. 
In our case, these Majorana fermions belong to the representation 
$(S_{tot},T_{tot})=(1/2,1/2)$ and are coupled to the magnetic field 
simultaneously. This leads to the result that the QCP moves to $h_c=2m$ 
instead of $m$. Beyond that value, all excitations become gapless and 
the correlation functions are algebraic decay. The situation is quite 
different from the two-leg spin ladder model by considering $\vec{T}_i$ 
in Eq. (\ref{KJ}) as the spin operator on the other chain. In that case, 
the magnetic field is only coupled to $\psi_+$ and the QCP is located at 
$h_c=m$. Beyond this value, $\psi_-$ is still massive and we expect that 
the behavior of correlation functions is similar to that of the usual spin 
ladders.

   What we are concerned is the case where $h>h_c$. The effective 
Hamiltonian describing the low energy excitations around the Fermi point 
is as the following:
\begin{equation}
  H = \sum_{a=\pm}\int dx \ \frac{v}{2} \ [(\partial_x\tilde{\theta}_a)^2
     +(\partial_x\tilde{\phi}_a)^2].
\label{wH3}
\end{equation}
Note that Eq. (\ref{wH3}) is in fact a theory describing two free 
massless fermions. We are now in a position to calculate the spin and 
orbital correlation functions. In terms of $\tilde{\phi}_a$ and 
$\tilde{\theta}_a$, the spin and orbital operators are expressed as 
follows\cite{Giam99}: 
\begin{eqnarray}
 S^z_n &=& \frac{M}{2}+\frac{a_0}{2\sqrt{\pi}}\partial_x(\tilde{\phi}_+
           +\tilde{\phi}_-)+\lambda_1(-1)^n\sin{[\frac{\pi M}{a_0}x+
           \sqrt{\pi}(\tilde{\phi}_++\tilde{\phi}_-)]}, \nonumber  \\
 S^+_n &=& \frac{1}{2\pi}\left[e^{i\frac{\pi M}{a_0}x}e^{i\sqrt{\pi}
           (\tilde{\phi}_++\tilde{\theta}_+)}e^{i\sqrt{\pi}(\tilde{\phi}_-
           +\tilde{\theta}_-)}+e^{-i\frac{\pi M}{a_0}x}e^{-i\sqrt{\pi}
           (\tilde{\phi}_+-\tilde{\theta}_+)}e^{-i\sqrt{\pi}(\tilde{\phi}_-
           -\tilde{\theta}_-)}\right] \nonumber  \\
   & & +\lambda_2(-1)^ne^{i\sqrt{\pi}(\tilde{\theta}_++\tilde{\theta}_-)}, 
           \nonumber \\
 T^z_n &=& \frac{a_0}{2\sqrt{\pi}}\partial_x(\tilde{\phi}_+-\tilde{\phi}_-)
           +\lambda_3(-1)^n\sin{[\sqrt{\pi}(\tilde{\phi}_+
           -\tilde{\phi}_-)]}, \nonumber  \\
 T^+_n &=& \frac{1}{2\pi}\left[e^{i\sqrt{\pi}(\tilde{\phi}_++
           \tilde{\theta}_+)}e^{-i\sqrt{\pi}(\tilde{\phi}_-
           +\tilde{\theta}_-)}+e^{-i\sqrt{\pi}(\tilde{\phi}_+
           -\tilde{\theta}_+)}e^{i\sqrt{\pi}(\tilde{\phi}_-
           -\tilde{\theta}_-)}\right]  \nonumber  \\
   & & +\lambda_3(-1)^ne^{i\sqrt{\pi}(\tilde{\theta}_+-\tilde{\theta}_-)}, 
\label{ds3}
\end{eqnarray}
where $x=na_0$ and $\lambda_i$ ($i=1,2,3$) are constants. Note that $S^z_n$ 
contains the $Q$ ($=\frac{\pi}{a_0}(1-M)$) term in contrast to the two-leg 
spin ladder where the term next to the uniform component is the $2Q$ 
one\cite{Giam99}. The reason is that in the spin ladder, the antisymmetrical 
mode $\phi_-$ is still massive even under the magnetic field and leads to the 
exponential decay of the $Q$ term. However, in our case, $\phi_-$ becomes 
massless when $h>h_c$. With the help of Eq. (\ref{ds3}), we get the 
correlators as the following:
\begin{eqnarray}
 \langle S^z_i(\tau)S^z_j(0) \rangle &=& \frac{M^2}{4}+\frac{a^2_0}
            {8\pi^2}\left(\frac{1}{z^2}+\frac{1}{\bar{z}^2}\right)+D_1\cos{
            (Qx)}\frac{a_0}{|z|}, \nonumber  \\
\langle S^+_i(\tau)S^-_j(0) \rangle &=& \frac{a_0^2}{4\pi^2|z|^2}\left(e^{
             i\frac{\pi M}{a_0}x}\frac{\bar{z}}{z}+e^{-i\frac{\pi M}{a_0}x}
             \frac{z}{\bar{z}}\right)+D_2(-1)^{\frac{x}{a_0}}\frac{a_0}
             {|z|}, \nonumber  \\
\langle T^x_i(\tau)T^x_j(0) \rangle &=& \langle T^y_i(\tau)T^y_j(0) 
             \rangle = \langle T^z_i(\tau)T^z_j(0) \rangle \nonumber  \\
    &=& \frac{a^2_0}{8\pi^2}\left(\frac{1}{z^2}+\frac{1}{\bar{z}^2}
             \right)+(-1)^{\frac{x}{a_0}}D_3\frac{a_0}{|z|}, 
\label{ds4}
\end{eqnarray}
where $x=a_0|i-j|$, $z=v\tau+ix$ and $D_i$ ($i=1,2,3$) are constants. Note 
that all correlators in Eq. (\ref{ds4}) behave like the spin correlation functions 
of an AF Heisenberg chain when $h \rightarrow h^+_c$ ($M\rightarrow 
0^+$). Low energy modes appear close to $q=0$ and $Q$ for $S^{zz}$ or 
$q=\frac{\pi M}{a_0}$ and $\frac{\pi}{a_0}$ for $S^{+-}$ while for spin 
ladders they are close to $q=0$ and $\frac{\pi M}{a_0}$ for $S^{zz}$ or 
$q=Q$ and $\frac{\pi}{a_0}$ for $S^{+-}$\cite{Giam99}. The orbital 
correlators are identical to the spin correlation functions of an AF Heisenberg 
spin chain up to some numerical prefactors for any finite $h$ ($>h_c$).   
 
\section{Conclusion}

  In this paper, we discuss the effects of a magnetic field on the one 
dimensional spin-orbital model with $J_1=J_2=J$. In the gapless phase, 
i.e. $J\leq K/4$, the spin and orbital correlation functions still exhibit 
power-law decay. The magnetic field manifests itself on the emergence of 
incommensurate soft modes with characteristic momenta depending on the 
magnetization. Furthermore, the exponents are also dependent of the 
magnetization though they have universal values at zero field. A distinctive 
feature in the case with orbital degeneracy is that the structures of 
orbital correlators are hardly influenced by the magnetic field. This is 
associated with the unbroken orbital SU(2) symmetry. 

   In the massive phase, i.e. $J>K/4$, we expect that there is a quantum 
phase transition induced by increasing the magnetic field. However, 
because of the different excitation spectrum at different range of $J/K$, the 
correlation functions exhibit distinct behavior between the region near the 
SU(4) point and the one with $K \ll J$.  In the former case, the magnetic 
field exerts a great influence on the $S^z=0$ sector such that the spectrum is 
completely different from that in the absence of the magnetic field. We 
propose that this change can be experimentally examined by studying the 
dynamical spin structure factors under a uniaxial pressure. In the latter case, 
the spin and orbital correlation functions become power-law decay when the 
magnetic field is greater than the critical value. Furthermore, the exponents 
take universal values which are identical to those of an AF Heisenberg 
chain. This is due to the fact that the underlying system is described by a 
free fermion theory. 

   Finally, we would like to mention the applicability of our results to the 
asymmetric case ($J_1\neq J_2$) around the SU(4) point. The massless 
and massive phases on the symmetric line now extend to large anisotropic 
regions\cite{Boulat99}. The phase boundary passes through the SU(4) 
point. The anisotropy affects not only the coupling constants in the 
Hamiltonian (\ref{workH}) but also the velocities of the spin and orbital 
sectors such that they become different. For the massless phase with 
$G_{2}=c(J_2-K/4)<0$ ($c>0$), the correlators under the magnetic field 
are similar to Eq. (\ref{stcf}) except that the velocity of $\xi^6$ is different 
from the one of the orbital sector. The symmetry of the fixed point 
Hamiltonian is in general U(1)$\times$SO(3)$\times$Z$_2$ instead of 
U(1)$\times$SO(4). In the massive phase with $G_2>0$, we still have an 
approximate U(1)$\times$SO(4) symmetry at low energy for $h>h_c$ and 
the correlators are the same as Eqs. (\ref{ds1}) and (\ref{ds2}). On the other 
hand, in the rest part of the phase diagram around the SU(4) point, it is 
possible that another phase transition of the KT type occurs on the $S^z=0$ 
sector when $h$ further increases\cite{Boulat99}. However, it is just a 
transition from the massive (our region II) to massless (our region I) phases. 
Therefore, the correlators are still of the forms we obtained in the 
corresponding regions. 

\acknowledgements
  Y.L. Lee is grateful to P. Lecheminant for enlightening discussions. The 
work of Y.L. Lee and Y.W. Lee is supported by National Science Council of 
R.O.C. under the Grant No. NSC89-2811-M-007-0015.
\appendix
\section{The spin and orbital operators}
   
   In this Appendix, we list the expressions of the WZNW fields 
$\vec{J}_{s(t)}$, $\vec{N}_{s(t)}$, and $\vec{n}_{s(t)}$ in the following:
\begin{eqnarray}
 \vec{J}_{t\nu} &=& \frac{i}{2} \ \vec{\xi}_{t\nu}\times \vec{\xi}_{t\nu},
                \ J^z_s = \frac{1}{\sqrt{\pi}}\partial_x \Phi_s, \nonumber \\
 J_{s+} &=& -\frac{i\eta_1}{\sqrt{\pi a_0}} \ \left\{\xi^6_R \exp{[-i 
                \sqrt{\pi}(\Phi_s- \Theta_s)]}+\xi^6_L \exp{[i\sqrt{\pi}
                (\Phi_s+ \Theta_s)]}\right\}, \nonumber \\
 N^z_s &=& A[\sin{(\sqrt{\pi}\Phi_s)}\mu_3\mu_4\mu_5\mu_6+i\cos{(
           \sqrt{\pi}\Phi_s)}\sigma_3\sigma_4\sigma_5\sigma_6], \nonumber  
           \\
 N_{s\pm} &=& A\eta_1\eta_3(\mu_3\mu_4\mu_5\sigma_6\pm \sigma_3\sigma_4
           \sigma_5\mu_6)\exp{(\pm i\sqrt{\pi}\Theta_s)}, \nonumber  \\
 N^z_t &=& A[\cos{(\sqrt{\pi}\Phi_s)}\sigma_3\sigma_4\mu_5\mu_6+i\sin{
           (\sqrt{\pi}\Phi_s)}\mu_3\mu_4\sigma_5\sigma_6], \nonumber  \\
 N_{t\pm} &=& -iA\eta_2\eta_3(\sigma_3\mu_4\pm i\mu_3\sigma_4)[\sin{(
           \sqrt{\pi}\Phi_s)}\mu_5\sigma_6\mp \cos{(\sqrt{\pi}\Phi_s)}
           \sigma_5\mu_6], \nonumber  \\
 \vec{n}_t &=& -iB \vec{\xi}_{tR}\times \vec{\xi}_{tL}, \ 
 n^z_s = \frac{B}{\pi a_0}\sin{(\sqrt{4\pi}\Phi_s)}, \nonumber  \\
 n_{s+} &=& \frac{iB\eta_1}{\sqrt{\pi a_0}} \ \left\{\xi^6_R\exp{[i
           \sqrt{\pi}(\Phi_s+\Theta_s)]}+\xi^6_L\exp{[-i\sqrt{\pi}
           (\Phi_s-\Theta_s)]}\right\}, 
\label{ST}
\end{eqnarray}
where $\nu =R,L$ and $\hat{O}_{\pm}=\hat{O}^x\pm i\hat{O}^y$ ($\hat{O}=
J_{s(t)}, N_{s(t)}, n_{s(t)}$). $A$ and $B$ are nonuniversal constants. 
$\vec{\xi}_{t\nu}=(i\xi^3_\nu , -i\xi^4_\nu , i\xi^5_\nu)$ is a vector 
under orbital SO(3) rotations. $\sigma_a$ and $\mu_a$ ($a=3,\cdots,6$) 
are order and disorder parameters of the corresponding Ising models. 

\section{Correlation functions in the region with $J>K/4$}
 
  Following the appendix of Ref.\cite{Giam99}, we found that in this case 
it is not necessary to include higher harmonics. When $h>h_c$, the spin 
operators become
\begin{eqnarray}
  S^z_i &=& \frac{M}{2}+\frac{a_0}{\sqrt{\pi}}\partial_x\tilde{\phi}+
                    \lambda(-1)^{\frac{x}{a_0}}\sin{(\frac{M}{a_0}x+
                    \sqrt{4\pi}\tilde{\phi})}, \nonumber  \\
  S^{\pm}_i &\sim& e^{i\frac{\pi}{2a_0}x}e^{\pm i\sqrt{\pi}
                    \tilde{\theta}}\cos{(\sqrt{2\pi}\Phi_{\mp})}+{\rm H.c.},
\label{jkcf}
\end{eqnarray}
where $\lambda$ is a constant. As for the orbital operators, the 
corresponding WZNW fields become 
\begin{eqnarray}
  J^z_t &=& \frac{1}{\sqrt{2\pi}}\partial_x(\Phi_++\Phi_-), \nonumber 
                    \\
  N^z_t &\sim& \cos{(\frac{\pi M}{2a_0}x+\sqrt{\pi}\tilde{\phi})}
                          \sin{[\sqrt{\frac{\pi}{2}}(\Phi_++\Phi_-)]}
                          \sin{[\sqrt{\frac{\pi}{2}}(\Theta_+-\Theta_-)]} 
                          \nonumber  \\
             & & +i\sin{(\frac{\pi M}{2a_0}x+\sqrt{\pi}\tilde{\phi})}
                          \cos{[\sqrt{\frac{\pi}{2}}(\Phi_++\Phi_-)]}
                          \cos{[\sqrt{\frac{\pi}{2}}(\Theta_+-\Theta_-)]}, 
                          \nonumber  \\
  n^z_t &\sim& \sin{[\sqrt{2\pi}(\Phi_++\Phi_-)]}.
\label{jc2}
\end{eqnarray}
The computations involving the $\tilde{\phi}$ sector are straightforward.  
To compute the longitudinal dynamical spin structure factor, we need the 
following integral:
\begin{eqnarray}
 \int^{\infty}_{-\infty}dx dt \ \frac{e^{i\omega t-iqx}}
      {(x+vt-i0^+)^{\alpha}(x-vt+i0^+)^{\beta}} &=& 
      \Theta(\omega -vq)\Theta(\omega +vq)\frac{2\pi^2e^{i\pi 
      (\alpha -\beta)/2}}{v\Gamma(\alpha)\Gamma(\beta)} 
      \nonumber  \\ 
  & & \times \left(\frac{2v}{\omega -vq}\right)^{1-\alpha}
          \left(\frac{2v}{\omega +vq}\right)^{1-\beta}.
\label{integ}
\end{eqnarray}

    The correlators involving $\Phi_{\pm}$ only can be calculated in terms 
of the exact results of form factors on the sine-Gordon 
model\cite{Kara78,Smir92,Luky97,Delf99}. The calculations can be 
partially simplified by noting that the Hamiltonian (\ref{deq}) is invariant 
under the exchange $\Phi_+ \leftrightarrow \Phi_-$ and $\Phi_+$ and 
$\Phi_-$ are decoupled. We now compute $T^{zz}(\omega, q\approx 0)$. 
(A similar calculation has been done in the context of a spin ladder 
model\cite{Allen99}.) With the help of Eq. (\ref{jc2}), it is
\begin{eqnarray}
  T^{zz}(\omega, q\approx 0) &=& \frac{2a^2_0}{\pi} \ {\rm Im} \lim_{
       i\omega \rightarrow \omega+i0^+}\int^{\infty}_{-\infty} 
       dx d\tau \ \langle \partial_x\Phi_+(\tau,x)\partial_x\Phi_+
       (0,0)\rangle e^{i\omega \tau-iqx} \nonumber  \\
   &=& 2a_0^2 \ {\rm Im} \ i\int^{\infty}_{-\infty}dx\int^{\infty}_{0}dt 
       \ e^{i(\omega+i0^+) t-iqx}\langle [j_0(t,x), j_0(0,0)]\rangle ,
\label{t1}       
\end{eqnarray}
where $j_0=\frac{1}{\sqrt{\pi}}\partial_x\Phi_+$ is the temporal component
of the current operator in the sine-Gordon model and we set $v_f=1$. In 
one dimension, energy and momentum can be parametrized by rapadity 
$\theta$ as $\epsilon =m\cosh{\theta}$ and $p=m\sinh{\theta}$ where $m$ 
is the mass of the sine-Gordon solitons. The resolution of the identity is 
given by
\begin{equation}
 1=\sum^{\infty}_{n=0}\sum_{\alpha_i}\int \frac{d\theta_1 \cdots d
   \theta_n}{(2\pi)^n n!}|\theta_n, \cdots ,\theta_1\rangle_{\alpha_n
   \cdots \alpha_1}{}^{\alpha_1 \cdots \alpha_n}\langle \theta_1, \cdots ,
   \theta_n|,
\label{t2}
\end{equation}
where $n$ is the number of particles. $\alpha_i=\frac{1}{2}$, $-\frac{1}
{2}$ for solitons and antisolitons, respectively. Inserting Eq. (\ref{t2}) 
into Eq. (\ref{t1}) and performing the integrations over space and time 
yields
\begin{eqnarray}
 T^{zz}(\omega ,q\approx 0) &=& -4\pi a^2_0 \ {\rm Im} \ 
       \sum^{\infty}_{n=0}\sum_{\alpha_i}\int \frac{d\theta_1 \cdots d
       \theta_n}{(2\pi)^n n!}|F^{j_0}_{\alpha_1 \cdots \alpha_n}(\theta_1, 
       \cdots ,\theta_n)|^2 \nonumber  \\
  & & \times \left[\frac{\delta (q-m\sum_j\sinh{\theta_j})}{\omega -m
      \sum_j\cosh{\theta_j}+i0^+}-\frac{\delta (q+m\sum_j\sinh{\theta_j})}
      {\omega +m\sum_j\cosh{\theta_j}+i0^+}\right],
\label{t3}
\end{eqnarray}
where
\begin{eqnarray}
  F^{j_0}_{\alpha_1 \cdots \alpha_n}(\theta_1,\cdots ,\theta_n) &\equiv& 
     \langle 0|j_0(0,0)|\theta_n, \cdots, \theta_1\rangle_{\alpha_n \cdots
     \alpha_1}  \nonumber 
\end{eqnarray}
is the sine-Gordon current form factor. (For an operator $\hat{O}(t,x)$, 
the form factor $F^{\hat{O}}_{\alpha_1 \cdots \alpha_n}(\theta_1,\cdots ,
\theta_n) \equiv \langle 0|\hat{O}(0,0)|\theta_n, \cdots, \theta_1\rangle_{
\alpha_n \cdots \alpha_1}$.) Note that an $n$-particle state 
only contributes to Eq. (\ref{t3}) when $s^2=\omega^2-q^2 \geq n^2m^2$ 
and $n$ must be an even integer. Thus, at low energy $s^2<16m^2$, only 
two-particle states contribute. The corresponding form factor\cite{Kara78}
is
\begin{eqnarray}
 F^{j_0}_{\frac{1}{2},-\frac{1}{2}}(\theta_1,\theta_2) &=& -2m\sinh{
          \left(\frac{\theta_1+\theta_2}{2}\right)} f(\theta_{12}), \nonumber  
\end{eqnarray}
where $\theta_{12}=\theta_1-\theta_2$ and 
\begin{equation}
 f(\theta)=\frac{i}{2\pi}\sinh{(\theta/2)}\times \exp{\left\{\int^{
      \infty}_0\frac{dx}{x} \ \frac{\sin^2{[(\theta -i\pi)x/2]}}{
      \sinh{(\pi x)}}[\tanh{(\pi x/2)}-1]\right\}}.
\label{t4}
\end{equation}
After performing the integrations over $\theta_i$, we obtain
\begin{equation} 
 T^{zz}(\omega ,q\approx 0)=8a^2_0\frac{m^2q^2|f(2\chi(s))|^2}
           {s^3\sqrt{s^2-4m^2}}sgn(\omega),
\label{t5}
\end{equation}
where $\chi(s)=\cosh^{-1}{(s/2m)}$ and $4m^2<s^2<16m^2$.

   Next, we compute $T^{zz}(\omega, q\approx \frac{\pi}{a_0})$. 
According to Ref.\cite{Zamo97}, $\langle \cos{(\frac{\beta}{2}\phi)}
\rangle ={\cal G}_{\beta/2}$ where 
\begin{eqnarray}
  {\cal G}_a &\equiv& \langle e^{ia\phi} \rangle \nonumber \\
    &=& \left[\frac{m\sqrt{\pi}\Gamma (4\pi/(8\pi -\beta^2))}{2\Gamma 
        (\beta^2/(16\pi -2\beta^2))}\right]^{\frac{a^2}{4\pi}} \nonumber  
        \\
    & & \times \exp{\left\{\int^{\infty}_0 \frac{dt}{t}\left[\frac{
        \sinh^2{(2a\beta t)}}{2\sinh{(\beta^2t)}\sinh{(8\pi t)}\cosh{
        ((8\pi -\beta^2)t)}}-\frac{a^2}{4\pi}e^{-16\pi t}\right]\right\}},
\label{ga}
\end{eqnarray}
and $\beta <\sqrt{8\pi}$ is the coupling constant of the sine-Gordon model. 
Therefore, the leading term of $\langle n^z_t(\tau ,x)n^z_t(0,0)\rangle 
\propto {\cal G}^2_{\beta/2}F(\sin{(\frac{\beta}{2}\Phi_+)})$ where 
$\beta^2=8\pi^-$. Then, 
\begin{eqnarray}
  T^{zz}(\omega, q\approx \frac{\pi}{a_0}) &\propto& {\rm Im} \ i 
     \int^{\infty}_{-\infty} dx \int^{\infty}_0 dt e^{i(\omega +i0^+)t-i
     (q-\frac{\pi}{a_0})x}\langle [\sin{(\frac{\beta}{2}\Phi_+)}(t,x),
     \sin{(\frac{\beta}{2}\Phi_+)}(0,0)]\rangle \nonumber  \\
  &\propto& {\rm Im} \ \sum^{\infty}_{n=0}\sum_{\alpha_i}\int \frac{d
     \theta_1 \cdots d\theta_n}{(2\pi)^n n!}|F^{\sin{(\frac{\beta}{2}
     \Phi_+)}}_{\alpha_1 \cdots \alpha_n}(\theta_1,\cdots ,\theta_n)|^2 
     \nonumber  \\
  & & \times \left[\frac{\delta (q-m\sum_j\sinh{\theta_j})}{\omega -m
      \sum_j\cosh{\theta_j}+i0^+}-\frac{\delta (q+m\sum_j\sinh{\theta_j})}
      {\omega +m\sum_j\cosh{\theta_j}+i0^+}\right].
\label{t6}
\end{eqnarray}
The leading contributions to Eq. (\ref{t6}) come from the two-particle 
states and the corresponding form factor\cite{Delf99} is
\begin{eqnarray}
 F^{\sin{(\frac{\beta}{2}\Phi_+)}}_{12}(\theta_{12}) &=& -F^{\sin{(\frac{
     \beta}{2}\Phi_+)}}_{21}(\theta_{12})\propto \frac{F_0(\theta_{12})}
     {\cosh{[\frac{\pi}{2\zeta}(\theta_{12}-i\pi)]}}, \nonumber
\end{eqnarray}
where $\zeta=\pi \beta^2/(8\pi -\beta^2)$ and 
\begin{equation}
 F_0(\theta)=-i\sinh{(\theta/2)}\times \exp{\left\{\int^{\infty}_0 \frac{dx}
    {x} \frac{\sinh{(\frac{\pi -\zeta}{2}x)}\sin^2{[\frac{x}{2}(\theta -
    i\pi)]}}{\sinh{(\pi x)}\sinh{(\frac{\zeta x}{2})}\cosh{(\frac{\pi x}{2})
    }}\right\}}.
\label{t7}
\end{equation}
Here the indices of the form factor $\alpha_i=1,2$ correspond to the neutral 
fermions which are related to the sine-Gordon solitons through $Z_{\pm}=
Z_1\pm iZ_2$ where $Z_{\pm}$ and $Z_{1,2}$ are, respectively, annihilation 
operators of solitons (anti-solitons) and neutral fermions\cite{Delf99}. 
Inserting Eq. (\ref{t7}) into Eq. (\ref{t6}) and performing the integrations 
over $\theta_i$, we get
\begin{equation}
 T^{zz}(\omega, q\approx \frac{\pi}{a_0}) \propto \frac{m^2}{u\sqrt{
     u^2-4m^2}}\times \frac{|F_0(2\chi (u))|^2}{\cos{(\frac{\pi^2}{\zeta})
     }+\cosh{[\frac{2\pi}{\zeta}\chi (u)]}}sgn(\omega),
\label{t8}
\end{equation}
where $4m^2 < u^2=\omega^2-(q-\pi/a_0)^2 <16m^2$. 
       


\begin{references}
\bibitem{Kugel82} K.I. Kugel and D.I. Khomskii, Usp. Fiz. Nauk {\bf 136}, 
                  621 (1982) [Sov. Phys. Usp. {\bf 25}, 231 (1982)].
\bibitem{Broh96} C. Broholm, G. Aeppli, S.-H. Lee, W. Bao, and J.F. 
                 DiTusa, J. Appl. Phys. {\bf 79}, 5023 (1996); W. Bao, 
                 C. Broholm, G. Aeppli, P. Dai, J.M. Honig, and 
                 P. Metcalf, Phys. Rev. Lett. {\bf 78}, 507 (1997).
\bibitem{Axtel97} E. Axtell, T. Ozawa, S. Kauzlarich, and R.R.P. Singh, 
                  J. Solid State Chem. {\bf 134}, 423 (1997).
\bibitem{Isob96} M. Isobe and Y. Ueda, J. Phys. Soc. Jpn. {\bf 65}, 1178 
                 (1996); Y. Fujii, H. Nakao, T. Yosihama, M. Nishi, K. 
                 Nakajima, K. Kakurai, M. Isobe, Y. Ueda, and H. Sawa, 
                 {\it ibid}. {\bf 66}, 326 (1997).  
\bibitem{YQLi98} Y.Q. Li, M. Ma, D.N. Shi, and F.C. Zhang, Phys. Rev. 
                 Lett. {\bf 81}, 3527 (1998).
\bibitem{Yama98} Y. Yamashita, N. Shibata, and K. Ueda, Phys. Rev. {\bf B 
                 58}, 9114 (1998).
\bibitem{Pati98} S.K. Pati, R.R.P. Singh, and D.I. Khomskii, Phys. Rev. 
                 Lett. {\bf 81}, 5406 (1998).
\bibitem{Frisch99} B. Frischmuth, F. Mila, and M. Troyer, Phys. Rev. 
                   Lett. {\bf 82}, 835 (1999).
\bibitem{Azaria99} P. Azaria, A.O. Gogolin, P. Lecheminant, and 
                   A.A. Nersesyan, Phys. Rev. Lett. {\bf 83}, 624 (1999).
\bibitem{YQLi99} Y.Q. Li, F.C. Zhang, M. Ma, and D.N. Shi, 
                 cond-mat/9902269.
\bibitem{Yama99} Y. Yamashita, N. Shibata, and K. Ueda, cond-mat/9905211.
\bibitem{Yamash} Y. Yamashita, N. Shibata, and K. Ueda, cond-mat/9908237.
\bibitem{Boulat99} P. Azaria, E. Boulat, and P. Lecheminant, cond-mat/9910218.
\bibitem{Suth75} B. Sutherland, Phys. Rev. {\bf B 12}, 3795 (1975).
\bibitem{Affleck86} I. Affleck, Nucl. Phys. {\bf B 265}, 409 (1986); 
                    {\it ibid}. {\bf 305}, 582 (1988).
\bibitem{Nerses97} A.A. Nersesyan and A.M. Tsvelik, Phys. Rev. Lett. 
                   {\bf 78}, 3939 (1997).
\bibitem{foot1} There is another scenario about the behavior of the 
                structure factor in the region with $K/4< J\leq K/2$ 
                suggested in Ref.\cite{Pati98}. It is that there is an 
                incommensurate peak which changes from $q=\pi/2$ to $q=
                \pi$ with increasing $J/K$. 
                 
\bibitem{Assa99} R. Assaraf, P. Azaria, M. Caffarel, and P. Lecheminant, 
                 Phys. Rev. {\bf B 60}, 2299 (1999).
\bibitem{foot2} The choice of the Klein factors does not affect our 
                results. In the present paper we choose them such that 
                $\eta_{2\uparrow}\eta_{2\downarrow}\eta_1\eta_3=1=-
                \eta_{1\uparrow}\eta_{1\downarrow}\eta_1\eta_3=\eta_{1
                \uparrow}\eta_{2\uparrow}\eta_2\eta_3=\eta_{1\downarrow}
                \eta_{2\downarrow}\eta_2\eta_3$. These are all what we 
                need.
\bibitem{Affleck89} I. Affleck, in {\it Fields, Strings, and  Critical 
                    Phenomena}, edited by E. Brezin and J. Zinn-Justin, 
                    (Elsevier Science, New York, 1989).
\bibitem{Giam99} T. Giamarchi and A.M. Tsvelik, Phys. Rev. {\bf B 59}, 
                              11398 (1999).
\bibitem{Shan78} R. Shankar and E. Witten, Nucl. Phys. {\bf B 141}, 329 
                 (1978).  
\bibitem{Witten78} E. Witten, Nucl. Phys. {\bf B 142}, 285 (1978).
\bibitem{Kara78} M. Karawski and P. Weisz, Nucl. Phys. {\bf B 139}, 445 
                 (1978).
\bibitem{Smir92} F.A. Smirnov, {\it Form Factors in Completely Integrable 
                 Models of Quantum Field Theory}, (World Scientific, 
                 Singapore, 1992).
\bibitem{Luky97} S. Lukyanov, Mod. Phys. Lett. {\bf A 12}, 2543 (1997).
\bibitem{Delf99} G. Delfino, Phys. Lett. {\bf B 450}, 196 (1999).
\bibitem{Zamo97} S. Lukyanov and A. Zamolodchikov, Nucl. Phys. {\bf B 493},
                 571 (1997).
\bibitem{Allen99} D. Allen, F.H.L. Essler, and A.A. Nersesyan, 
                  cond-mat/9907303.  
\end{references}
\end{document}